\documentclass[openacc]{rsproca_new}

\graphicspath{{fig/}}

\usepackage{cleveref}

\titlehead{Research}

\begin{document}

\title{Identification of Empirical Constitutive Models for Age-Hardenable
Aluminium Alloy and High-Chromium Martensitic Steel Using Symbolic Regression}

\author{
E. Kabliman$^{1,2}$, G. Kronberger$^{3}$}

\address{$^{1}$University of Bremen, Faculty 04: Production Engineering – Mechanical Engineering and Process Engineering, Badgasteinerstrasse 1, 28359 Bremen, Germany\\
$^{2}$Leibniz Institute for Materials Engineering – IWT, Badgasteinerstrasse 3, 28359 Bremen, Germany\\
$^{3}$University of Applied Sciences Upper Austria, Heuristic and Evolutionary Algorithms Laboratory (HEAL), Softwarepark 11, 4232 Hagenberg\\
}

\subject{Constitutive modelling, Material Science, Symbolic Regression, Evolutionary Algorithms}


\corres{Evgeniya Kabliman\\
\email{kabliman@uni-bremen.de}}

\begin{abstract}
Process-structure-property relationships are fundamental in materials science and engineering and are key to the development of new and improved materials. Symbolic regression serves as a powerful tool for uncovering mathematical models that describe these relationships. It can automatically generate equations to predict material behaviour under specific manufacturing conditions and optimize performance characteristics such as strength and elasticity. 

The present work illustrates how symbolic regression can derive constitutive models that describe the behaviour of various metallic alloys during plastic deformation. Constitutive modelling is a mathematical framework for understanding the relationship between stress and strain in materials under different loading conditions. In this study, two materials (age-hardenable aluminium alloy and high-chromium martensitic steel) and two different testing methods (compression and tension) are considered to obtain the required stress-strain data. The results highlight the benefits of using symbolic regression while also discussing potential challenges.
\end{abstract}

\begin{fmtext}
\end{fmtext}

\maketitle


\section{Introduction}

In materials science and engineering, understanding the correlations between processes, structures, and properties is fundamental to developing new materials and processes. As artificial intelligence and machine learning gain prominence in this field, it is essential to focus on methods that provide formal descriptions of these relationships. In this context, mathematical expressions are valuable because they enable a direct examination of relationships and serve as a universal language in science. In addition, incorporating mathematical expressions into numerical simulation frameworks, which are increasingly used in place of physical experiments, is relatively straightforward. Therefore, expressing models as mathematical equations can have significant scientific and practical implications. Symbolic regression can automatically generate these equations, making its application in materials science and engineering highly valuable.

In manufacturing, materials undergo various transformations under applied conditions, such as temperature and pressure. Optimising manufacturing parameters typically requires extensive trial-and-error studies, which are often not feasible on an industrial scale. This work focuses on the deformation of metals. Metal forming is a manufacturing process in which a metal workpiece is reshaped through plastic deformation without the addition or removal of material. This reshaping is usually achieved by applying compressive, tensile, or shear forces using tools such as dies, rollers, or hammers. To investigate material behaviour during deformation, lab-scale deformation tests are commonly performed. The choice of a testing method depends on the specific manufacturing process or application. For example, during the extrusion process or in crash simulations, high deformation rates are expected and can be evaluated using compression tests. In contrast, various components of power plants operate under creep conditions, which can be tested by applying very small strain rates. The test results (data) can differ depending on factors such as the deformation mode, the deformation rate, temperature, and the material itself. Therefore, it is essential to develop a modelling approach that can accurately predict material responses to applied deformation, regardless of the mentioned factors. 

The approach proposed in this work utilizes symbolic regression~\cite{Koza1992,Langdon2002,Schmidt2009,Kronberger2024}, which can be effectively solved using genetic programming. The equations obtained through symbolic regression provide the possibility for interpretation, potentially leading to new and improved insights into the system being studied. For example, it can reveal the influence of individual alloying elements or processing parameters such as temperature. This approach facilitates the design of new materials with desired properties and optimization of manufacturing processes without relying on predefined models or assumptions. 

Symbolic regression using genetic programming was successfully used to generate data-driven constitutive models earlier. For example, Version et al.~\cite{Versino2017} used this approach to describe stress-strain curves for metallic materials such as copper, while porous materials were studied in \cite{Bomarito2021}. These works highlight the potential of SR for interpretation and model analysis, in particular analysis of asymptotic behaviour. Furthermore, it has been shown that learning short equations from data can lead to improved generalization error even when extrapolating from the region covered by training examples~\cite{LearningEquations2018} and several techniques have been shown to improve symbolic regression extrapolation~\cite{OrderOfNonlinearity2009,KronbergerShapeConstrained}.
Thus, through symbolic regression, we aim to meet all the requirements outlined in the present work.

\begin{itemize}
    \item No prior knowledge is known about the system
    \item Searched model has a form of mathematical expressions
    \item Model transferability is required
    \item Available data is limited
    \item Prediction for deformation parameters not included in the training set
\end{itemize} 

Using the proposed approach, we will derive constitutive models, which serve as a central mathematical framework to understand the relationship between stress and strain in materials under various loading conditions. Constitutive models are at the core of numerical simulations of deformation processes such as rolling, extrusion, and forging, which makes them crucial for materials engineers. Compared to our previous work~\cite{10.1063/1.5112747,KRONBERGER2022100080,KABLIMAN2021100052}, we will use a new implementation of symbolic regression and account for potential variations in material behaviour depending on the deformation mode such as compression and tension. We will derive the models for both cases using the same approach and discuss the similarities and differences in applying symbolic regression to derive the constitutive models. Furthermore, we will consider two different materials, an age-hardenable aluminium alloy and a high-chromium martensitic steel. 

\section{Methods}

\subsection{Data acquisition}
The experimental data used in the present study were previously published by Kabliman et al.~\cite{KABLIMAN2021100052} (compression testing of the age-hardenable aluminium alloy) and Eisenträger et al.~\cite{Eisentraeger2017} (tensile testing of the high-chromium martensitic steel). The deformation temperature and the deformation rate were varied in both series of tests. The stress values were automatically calculated from force-displacement measurements, and the data was prepared accordingly. When analyzing the final stress-strain curves, a significant difference related to the deformation mode was observed. In compression testing, the stress-strain curve demonstrated a plateau after reaching its maximum value, while the tensile stress-strain curve showed a sharp decrease due to strain softening and necking. For model development, this indicated a higher level of complexity that needed to be captured. The further issue was that the transition from the linear elastic to non-linear plastic regime (yield point) was not well defined as well as the onset of necking. In general, the model developed in the present work had to capture the following characteristics:

\begin{itemize}
    \item linear elastic regime 
    \item strain hardening regime (non-linear behaviour)
    \item strain softening and necking regimes (non-linear behaviour)
\end{itemize}

\textbf{Compression test}. A series of hot compression tests was conducted using the deformation and quenching dilatometer DIL805A/D from TA Instruments. The specimens were made of the AA6082 aluminium alloy (main alloying elements: Mg and Si) and had a cylindrical form. They were compressed at the seven different deformation temperatures $T~=~\{350, 375, 400, 425$, $450, 475, 500\}$~\textdegree C and five different deformation rates $\dot\varphi = \{0.001, 0.01$, $0.1, 1, 10\} ~s^{-1}$ up to a total strain $\varphi$ of $0.7$. Two tests were performed for each temperature and strain rate, and the average was calculated to reduce data noise. This aggregation also reduced the uncertainties possible due to variations in the microstructure of the individual samples. Additionally, the aggregation simplifies the symbolic regression process. The drawback however is that information about sample uncertainty is lost. This information could be helpful for the modelling process in particular for model selection. We used the aggregate data nevertheless since only these data was available.

\textbf{Tensile test}. A series of high-temperature uniaxial tensile tests was performed on an electromechanical testing frame, CERMAC CSR50, with a tubular
resistance furnace. Two linear variable differential transformers have been mounted to the collars of
the specimen to record the displacement such that elongation is measured. The shouldered round specimens made of the high-chromium X20CrMoV12-1 steel were tested at six different temperatures $T = \{673, 723, 773, 823, 873, 923\}$ K (or $T = \{400, 450, 500, 550, 600, 650\}$ \textdegree C) and three different deformation rates $\dot\varepsilon = \{5\cdot10^{-5}, 10^{-4}, 10^{-3}\} ~s^{-1}$. The testing was performed until rupture. Both the engineering and true stress-strain curves exhibited significant softening behaviour due to possible material softening and necking effects. Because true stress-strain data was only available for the plastic region, the engineering stress-strain data was utilized for modelling in this study.

Finally, we kept the original notations for the stress and strain as given in Table \ref{tab:notations}. Herein, $F$ is the measured force, $L_0$ and $d_0$ are the initial length and diameter of a specimen, $L$ is the actual length of the specimen, while the actual diameter of a specimen could not be measured either in compression or tensile testing. Further experimental details can be found in~\cite{KABLIMAN2021100052} and~\cite{Eisentraeger2017}, correspondingly.

\begin{table}[!h]
\caption{Measured and derived variables from compression and tensile testing.}
\label{tab:notations}
\begin{tabular}{llll}
\hline
Variable &Compression test~\cite{KABLIMAN2021100052} &Tensile test~\cite{Eisentraeger2017} &Formula \\
\hline
Initial cross section &$A_0$ &$A_0$ &$\pi d_0^2/4$ \\
Actual cross section &$A$ &$A$ &$A_0 L_0/L$ \\
Engineering stress &- &$P$ &$F/A_0$ \\
Engineering strain &- &$\varepsilon$ &$(L-L_0)/L_0$ \\
True stress &kf &$\sigma$ &$F/A$ \\
True strain &$\varphi$ &$H$ &$\ln(L/L_0)$ \\
Deformation rate &$\dot \varphi$ &$\dot \varepsilon$ &$d\varphi/dt$; $d\varepsilon/dt$ \\\hline
\end{tabular}
\vspace*{-4pt}
\end{table}

\subsection{Data resampling}
The number of measurements collected in each test depends on the strain rate, that is, there are fewer measurements for higher strain rates because the test is much shorter. To produce a training dataset with an equal number of data points for each strain rate, we resampled the observed stress-strain curves in a regular interval of strain using linear interpolation. Through this procedure, we generated training sets with interpolated data points for each temperature and strain-rate combination (1200 points for the full stress-strain curves and 400 points for the curves cut off at maximum stress). The same linear interpolation procedure was used for both cases: compression testing of the AA6082 aluminium alloy and the tensile testing of the X20CrMoV12-1 martensitic steel. Prior explorative visualization uncovered three outlier stress-strain curves for the X20CrMoV12-1 dataset that could be visually distinguished, namely \{$\dot\varepsilon=5\cdot10^{-5}, T=673$ K\}, \{$\dot\varepsilon=1\cdot10^{-3}, T=673$ K\} and \{$\dot\varepsilon=1\cdot10^{-3}, T=773$ K\}. We completely removed these curves from the training set because we suspect an inconsistency in calibration or preprocessing. We still evaluate and visualize the predictions of the identified model on the full dataset in the results section.

\subsection{Symbolic regression and genetic programming}
Symbolic regression is a supervised learning task, where in the most simplified form the goal is to find an expression for the parametric regression function $f$ mapping inputs $x$ to outputs $y=f(x,\theta)$ with minimal errors. The specific challenge is that both the expression structure of the regression function $f$ and the coefficients $\theta$ have to be fitted to the dataset. The term symbolic regression was introduced by Koza~\cite{Koza1992} in the context of genetic programming (GP) as one of multiple similar tasks that can be solved with GP. 

In our previous work\cite{10.1063/1.5112747,KRONBERGER2022100080,KABLIMAN2021100052}, we used the software implementation of a tree-based genetic programming for symbolic regression called HeuristicLab~\cite{Wagner2014}. For the new experiments, we use pyoperon (version 0.5.0) which is a Python wrapper for Operon~\cite{Burlacu2020} to produce all models. Operon is the highly tuned implementation of tree-based genetic programming which uses the multi-objective algorithm to produce a set of expressions that are Pareto optimal in accuracy and complexity. \Cref{tab:pyoperon-parameters} shows the parameter settings that we used for all experiments. The objective of the algorithm is to minimize the mean of squared errors (MSE) and expression length.

\begin{table}
\caption{Pyoperon parameter settings used in all experiments. aq$(a,b) = a (1 + b^2)^{-1/2}$, $\div$ is allowed with a single argument $\text{div}(x) \equiv x^{-1}$}
\label{tab:pyoperon-parameters}
\begin{tabular}{ll}
\hline
  Parameter & Value \\
  \hline
  Population size & 1000 \\
  Pool size & 1000 \\
  Initialization & Balanced tree creator~\cite{Burlacu2020} \\
  Algorithm & NSGA-II~\cite{Deb2002} (keep best, replace worst) \\
  Selection & Tournament with crowding distance \\
            & group size 3 \\
  Generations & 100\\
  Mutation probability & 25 \% \\
  Crossover rate & 100 \% \\
  Internal crossover probability & 90 \% \\
  Maximum tree size & 100 nodes \\
  Local optimization & 10 Levenberg-Marquardt iterations (see e.g.~\cite{Fletcher2000})\\
  Objective 1 & Mean of squared errors (without linear scaling)\\
  Objective 2 & Expression tree size (length) \\
  Function set & param, var, $+, -, \times, \div, \exp, \log, x^2, \sqrt{x}, \tanh, $ aq \\\hline
\end{tabular}
\end{table}

We produced expressions for both materials and for two different scenarios. In the first scenario, we tried to find expressions that fit the full stress-strain curve. In the second scenario, we used only the observations up to the point of maximum stress. The same model building process is used for both materials and both scenarios. 
In each case, the input was a preprocessed data file, and then we used 8-fold cross-validation to determine the preferable expression length $k$. Afterwards, we ran GP again, but using the full training set, and selected the length $k$ expression from the Pareto-front. Finally, the expression was simplified, and the values of the simplified model parameters were optimized from the values identified by GP using nonlinear least-squares (to make sure that the values are a local optimum).

\subsection{Cross-validation and model selection}
Pyoperon produces a set of the Pareto-optimal expressions from the final population. This set contains the shortest and most accurate expressions visited by the algorithm over all generations. The selection of the final model from the  Pareto-front is critical, because it may contain overfit individuals which would exhibit high prediction errors. We used a form of cross-validation to identify the preferable expression lengths for each scenario individually. 

For 8-fold cross-validation, we assigned data from each combination of strain rate and temperature values either completely to the training or to the CV hold-out sets (the fold assignments are shown in the supplement). The result of the CV was 8 sets of Pareto fronts. We evaluated each expression on the CV hold-out set to get 8 training MSE and 8 hold-out MSE values for each expression length. The CV error for each expression length was calculated as the mean over the hold-out MSE values. We selected the minimal length $k$ for which the CV error is within one standard error of the best CV error.

\begin{figure}[ht]
\centering
\input{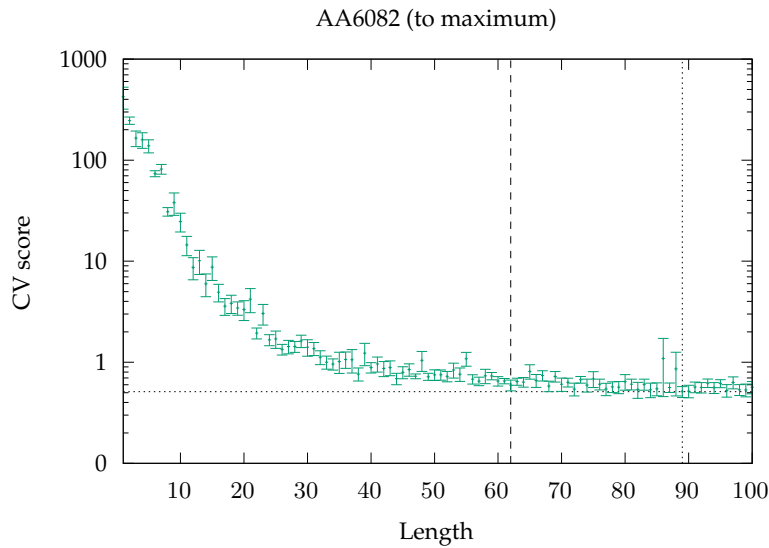}

\caption{Examples for the selection of best expression length using cross-validation for the AA6082 dataset cut off at the maximum stress. We collect $k$ Pareto-fronts from $k$-fold CV and calculate the CV-score as the mean of $k$ MSE values on the hold-out fold. The confidence bounds show the standard error of the CV-score (details in the supplement). Here, the best CV score is found for expressions with length of 89, but we select the shorter length of 62 symbols.} 
\label{fig:cv-scores}
\end{figure}

\Cref{fig:cv-scores} shows an example result for the AA6082 dataset with measurements up to the maximum stress value. For this dataset, the best CV-score was found for expressions of length 89. However using the one-sigma rule~\cite{Hastie2001}, we select the shorter length limit of 62 symbols, because that CV-score is within one sigma of the best one. This plot does not show signs of overfitting for models up to  length of 100. For the X20CrMoV12-1 dataset, we observed that GP started to overfit already for small lengths of approximately 20--40 symbols. The plots for all datasets are shown in the supplement. 

Within the evolutionary loop, we used memetic optimization of coefficients via non-linear least-squares using the Levenberg-Marquardt algorithm for all evaluated solution candidates, as described in~\cite{Kommenda2020} but disabled linear scaling of regression functions.

\section{Results \& Discussion}

\begin{figure}
\input{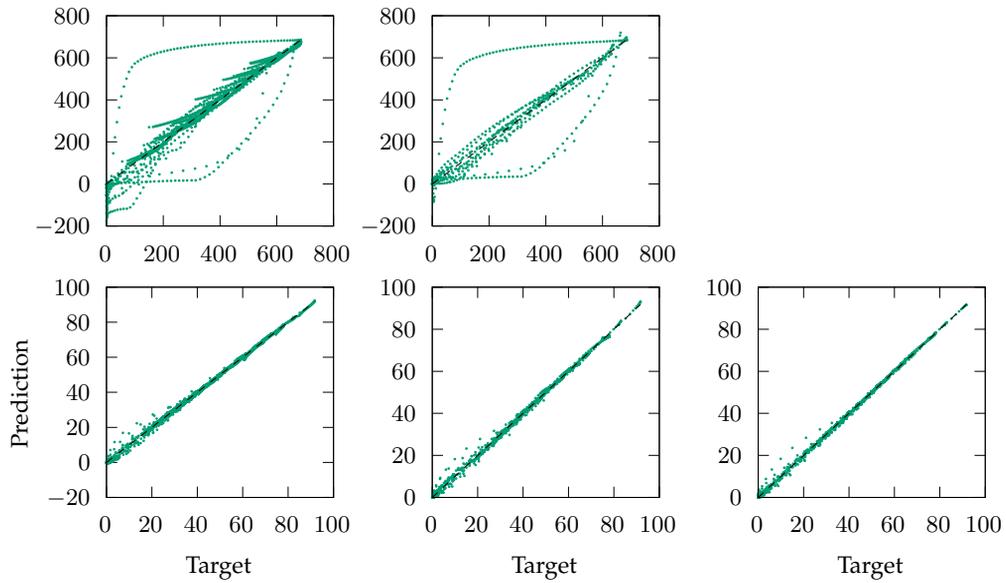}
\caption{Scatter plots of predicted over measured stresses. Top row: X20CrMoV12-1 (left: full curve, right: to maximum stress). Bottom row: AA6082 (left: full curve, middle and right: to maximum stress. The three outlier curves for X20CrMoV12-1, that were not used in training are clearly visible.  }\label{fig:scatterplots}
\end{figure}

\Cref{fig:scatterplots} shows the scatter plots of predicted over measured stress values for all models. 
In these plots, we show the measurements from all tests including those that were not used in the training phase.
For the AA6082 dataset, the prediction is more accurate than for X20CrMoV12-1. The high predictive accuracy of the model, in particular for the maximum stress value, is notable. The three outlier curves in the X20CrMoV12-1 are are visible as outliers in both scatter plots.

\begin{figure}
\input{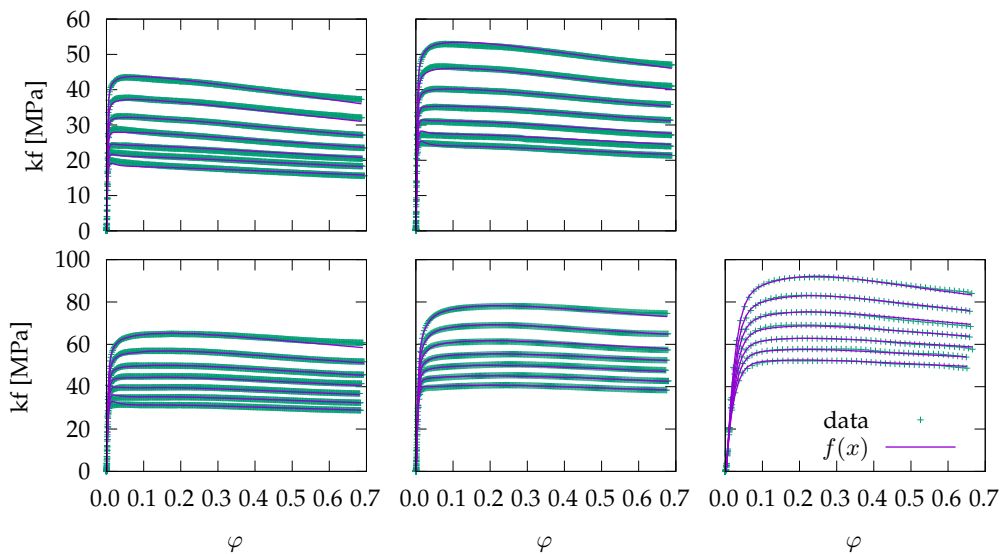}
\caption{Line charts of predictions and measurements of kf over $\varphi$ for the AA6082 model. The predictions closely align with the measurements. Temperatures: 350, 375, 400, 420, 450, 475, 500 \textdegree C from highest to lowest stress. $\dot\varphi: 0.001, 0.01, 0.1, 1, 10~s^{-1}$ from left to right and top to bottom.}
\label{fig:linechart-alu}
\end{figure}

\begin{figure}
\input{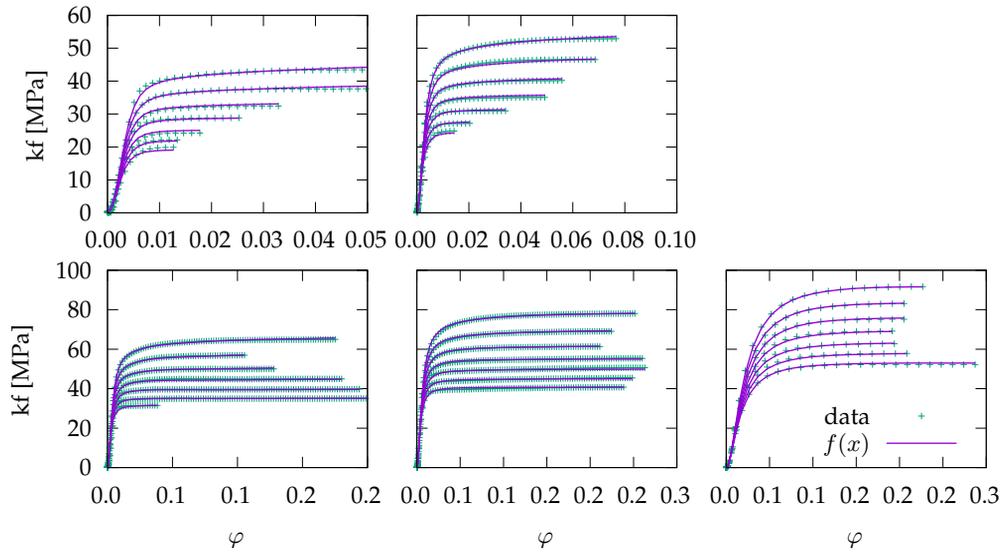}
\caption{Line charts of predictions and measurements of kf over $\varphi$ for the AA6082 model up to the maximum stress point. The predictions closely align with the measurements. Temperatures: 350, 375, 400, 420, 450, 475, 500 \textdegree C from highest to lowest stress. $\dot\varphi: 0.001, 0.01, 0.1, 1, 10~s^{-1}$ from left to right and top to bottom.}\label{fig:linechart-alu-tomax}
\end{figure}

\Cref{fig:linechart-alu} shows the evaluation results of \Cref{eq:alu-full} (AA6082, full curve) and  \Cref{fig:linechart-alu-tomax}  shows the evaluation results of \Cref{eq:alu-tomax-long} (AA6082, to maximum stress) as line charts compared to measurement data. Overall the prediction of the models for both scenarios are highly accurate with MSE$=0.22$ MPa$^2$ (full curve) and $0.32$ MPa$^2$ (to maximum stress) on the upsampled training data.

\begin{figure}
\input{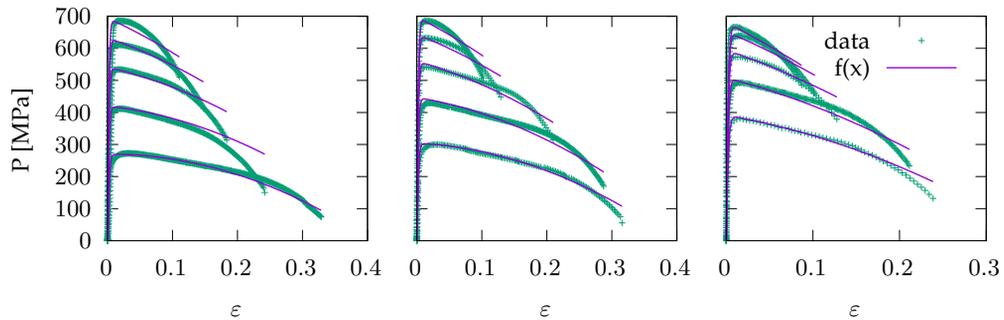}
\caption{Line charts of predictions and measurements of P over $\varepsilon$ for the X20CrMoV12-1 model. left: $\dot\varepsilon=5\cdot10^{-5}~s^{-1}$, middle: $\dot\varepsilon=1\cdot10^{-4}~s^{-1}$, right: $\dot\varepsilon=1\cdot10^{-3}~s^{-1}$. }\label{fig:linechart-steel-abb2}
\end{figure}

\Cref{fig:linechart-steel-abb2} shows the evaluation results of \Cref{eq:steel-abb2-full} (X20CrMoV12-1, full curve) and  \Cref{fig:linechart-alu-tomax}  shows the evaluation results of \Cref{eq:steel-abb2-tomax} (X20CrMoV12-1, to maximum stress) as line charts compared to measurement data. The model for the full curve is not able to predict the non-linearly reducing force at the end of the tests.

Both models accurately predict the maximum stress values. Notably, the X20CrMoV12-1 model demonstrates the ability to predict maximum stress even for outlier curves that were not included in the training data. The only exception is for the largest strain rate and lowest temperature, where the predictions tend to be overly high.

\begin{figure}
\input{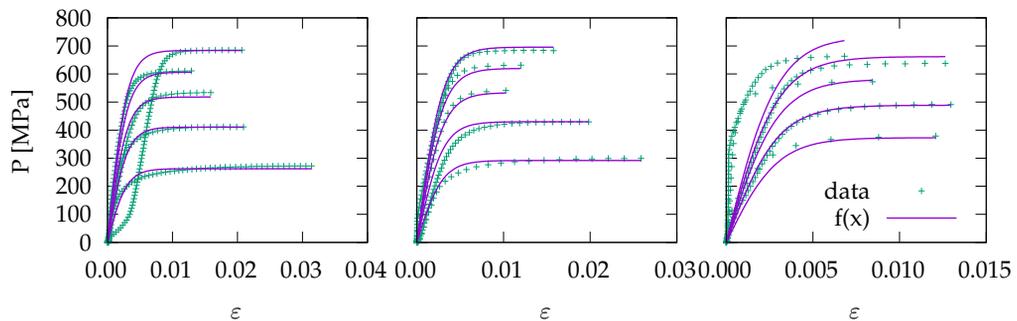}
\caption{Line charts of predictions and measurements of P over $\varepsilon$ for the X20CrMoV12-1 model up to the maximum stress. 
Although the curve's shape is affected by three outliers, the maximum stress value is predicted accurately, except for the curve with the highest strain rate and lowest temperature.
left: $\dot\varepsilon=5\cdot10^{-5}~s^{-1}$, middle: $\dot\varepsilon=1\cdot10^{-4}~s^{-1}$, right: $\dot\varepsilon=1\cdot10^{-3}~s^{-1}$.}\label{fig:linechart-steel-abb2-tomax}
\end{figure}

The expressions produced by pyoperon are shown below after simplification using SymPy. 
Each equation has a separate coefficient vector $\theta$ which is shown in \Cref{tab:coefficients}. 
\Cref{eq:alu-full} is the identified expression for the full AA6082 curve, \Cref{eq:alu-tomax-long} is the expression for the curve up to maximum stress.  
\Cref{eq:steel-abb2-full} is for the full X20CrMoV12-1 curve, and \Cref{eq:steel-abb2-tomax} is only valid up to the maximum stress.

The most striking difference is the length or complexity of the expressions for AA6082 and X20CrMoV12-1. We had less data for the high-chromium martensitic steel and found that larger models lead to overfitting, which is an indication that the data could be less accurate than the data for AA6082. We were able to produce much longer and much more accurate expressions without running into overfitting issues for the aluminum alloy. It is interesting to note that the expressions contain the logarithmic and hyperbolic functions as well as square root typically found in physics-based constitutive models.
The pipeline of systematically assigning all measurements from a single test to a single fold for cross-validation leads to an appropriate expression length for model selection, even though we do not have information about measurement uncertainties for both datasets.

\begin{align}
\hat{\text{kf}}_\text{full} &= \tanh\left( \theta_{1} \, \phi + \theta_{2} \, \log_{10}(\dot\phi) \right) \nonumber \\
    & \qquad \left(\theta_{3} \, \log_{10}(\dot\phi) + \left( \theta_{4} + \theta_{5} \, T \right) \, \tanh\left( \theta_{6} \, \phi + \theta_{7} \, \log_{10}(\dot\phi) \right)\right) \, \mathrm{aq}\left( t_1,\, t_2 \right) \label{eq:alu-full} \\
t_1 &= \theta_{8} \, \phi + \theta_{9} \, \log_{10}(\dot\phi) +  \theta_{10}^{\log_{10}(\dot\phi)} \, \theta_{11} \, T + \left( \theta_{12} + \theta_{13} \, T + \theta_{14} \, \dot\phi \right) \, \nonumber \\
   & \qquad \frac{\theta_{15} + \theta_{16} \, T + \left( \theta_{17} + \theta_{18} \, T \right) \, \tanh\left( \theta_{19} \, \phi + \theta_{20} \, \log_{10}(\dot\phi) \right)}{\sqrt{(1 + \theta_{21} ^{\log_{10}(\dot\phi)})(1 + \theta_{22} \, \left( \phi \right)^{-2})}} \nonumber \\
t_2 &= \theta_{23} + \theta_{24} \, T + \theta_{25} \, \log_{10}(\dot\phi) + \left( \theta_{26} + \theta_{27} \, T \right) \, \tanh\left( \theta_{28} \, \phi + \theta_{29} \, \log_{10}(\dot\phi) \right) \nonumber
\end{align}


\begin{align}
\hat{\text{kf}}_\text{max} &= \left( \theta_{1} ^{\log_{10}(\dot\phi)} \, \mathrm{aq}\left(t_{1} ,\, \theta_{14}^{\log_{10}(\dot\phi)} \, \mathrm{aq}\left( \theta_{15} + \theta_{16} \, T + \theta_{17} \, \dot\phi,\,  t_{2} \right) \right) \right)^{2}  \label{eq:alu-tomax-long} \\
t_{1} &= \theta_{2} + \theta_{3} \, T + \theta_{4} \, \log_{10}(\dot\phi) + \theta_{5} ^{\log_{10}(\dot\phi)} + \left( \theta_{6} + \theta_{7} \, T\right)^{2} \nonumber \\
     & \qquad \left(1 + \frac{\left( \theta_{8} + \theta_{9} \, \mathrm{aq}\left( \phi,\, \phi \,  \left(\theta_{10}\right)^{\log_{10}(\dot\phi)} \, \left( \theta_{11} + \theta_{12} \, T \right) \right) \right)^{2}}{1 + \theta_{13} \, \phi}\right)^{-1}  \nonumber \\
t_{2} &= \theta_{18} \, \tanh\left( \theta_{19} \, \phi \right) + \left(\theta_{20} + \theta_{21} \, \log_{10}(\dot\phi)\right)\left(1 + \frac{\left( \theta_{22} + \theta_{23} \, \dot\phi \right)^{2}}{\left( 1 + \theta_{24} \, \phi^{2} \right) \, \left( 1 + \left( \theta_{25} + \theta_{26} \, T \right)^{2} \right)}\right)^{-1/2} \nonumber
\end{align}

\begin{align}
\hat{P}_1 & = \tanh\left( \theta_{1} \, \varepsilon \right) \, \left( \theta_{2} \, T + \theta_{3} \, \varepsilon - (\theta_{4} \,  T) ^{2} - \left( \theta_{5} + \theta_{6} \, T + \theta_{7} \, \varepsilon + \theta_{8} \, \log_{10}(\dot\varepsilon) \right)^{2} \right) \label{eq:steel-abb2-full} \\
\hat{P}_2 & = \tanh\left( \theta_{1} \,  \varepsilon \right) \, \sqrt{\left|\theta_{2} + \theta_{3} \, T + \theta_{4} \, \log_{10}(\dot\varepsilon)\right|} \label{eq:steel-abb2-tomax}
\end{align}

\begin{table}
\caption{Coefficients for \Cref{eq:alu-full},  \Cref{eq:alu-tomax-long}, \Cref{eq:steel-abb2-full}, and \Cref{eq:steel-abb2-tomax}} \label{tab:coefficients}
\begin{tabular}{lcl}
\hline 
    Model & MSE & $\theta$  \\
          & train & \\
\hline    
    \cref{eq:alu-full} & $0.22$ & $[144.21, -0.30106, -17.754, -427.64, 0.50845, 370.89, -0.018056, 695.2,$ \\
     & & $20.061, 1.1864, -10.347, -203.97, 0.48387, -33.086, 2.6014, -0.063021,$ \\ 
     & & $-8.2764, 0.075567, 31.606, -0.18027, 4.7775, 745.51, -11892.0, 68.94, $\\
     & & $369.5, -2751.3, 2.9252, 6.2615, -0.28459]$ \\
    \cref{eq:alu-tomax-long} & $0.32$ & $[1.0109, 11.651, -0.012533, 0.38822, 1.4853, 2.9489, -0.0059347, 66.75,$\\
    & & $381.68, -2.3796\cdot10^{-11}, 132.71, -0.31013, 145030.0, 2.0142, 114.69, $\\
    & & $-0.13985, 6.4259, -2688.9, 3.9397, -19.082, -31.519, -1845.9, -49.51, $\\
    & & $3.4381\cdot10^7, -32.823, -0.0071516]$ \\
    \cref{eq:steel-abb2-full} & $233.8$ & $\left[347.07, 2.8128, -1061.9, 0.051378, 56.027, -0.056748, 40.733, 4.5755\right]$\\
    \cref{eq:steel-abb2-tomax} & $352.4$ & $\left[338.72, 2.042\cdot10^6, -1994.3, 53974.3\right]$ \\\hline
\end{tabular}
\end{table}

\section{Conclusion}
This study demonstrated the use of symbolic regression to derive the constitutive laws that describe how materials respond to applied deformation. We investigated two different materials: the age-hardenable aluminium alloy (AA6082) and the high-chromium martensitic steel (X20CrMoV12-1), testing them under two deformation regimes—compression and tension. By utilizing the corresponding stress-strain data, we successfully derived the constitutive equations for both materials and conditions. Notably, we applied the same model building approach and algorithm hyperparameters for both materials and testing scenarios, which ensured consistency and transferability in the studied examples, e.g. to other types of materials. The model building pipeline based on multi-objective genetic programming and systematic cross-validation could be used for both materials and testing scenarios to successfully produce accurate models with appropriate model length that did not show signs of overfitting. 

The findings highlight the common challenges and advantages associated with using symbolic regression. Similar to its applications in other scientific fields, symbolic regression effectively captured the complex relationships between applied deformation and material response. However, we discovered that even slight variations in the data can significantly increase the model complexity. Therefore, it is crucial to provide high-quality data including quantified measurement uncertainties, which can be difficult to obtain due to inherent heterogeneity in material microstructures, particularly in metallic alloys. Additionally, only a limited amount of data is often available because of experimental constraints, which can hinder the extrapolation capabilities of symbolic regression. One potential solution is to incorporate domain knowledge by applying prior constraints and integrating fundamental physics laws. Another approach is to use data from individual trials rather than relying on averaged values for the same set of testing parameters. Overall, we conclude that symbolic regression is a powerful tool for addressing challenges in materials science and engineering.

\vskip6pt

\enlargethispage{20pt}

\ack{E.K. acknowledges Dr. Johanna Eisenträger from Leibniz Institute for Materials Engineering - IWT for providing the mechanical testing data on X20CrMoV12-1 steel. 
G.K. acknowledges the Austrian Federal Ministry for Climate
Action, Environment, Energy, Mobility, Innovation and Technology,
the Federal Ministry for Labour and Economy, and the regional government of Upper Austria within the COMET project ProMetHeus
(904919) supported by the Austrian Research Promotion Agency
(FFG)}




\vskip2pc

\bibliographystyle{RS} 

\bibliography{symreg} 

\end{document}